\documentclass[prd,superscriptaddress,nofootinbib,amsmath,amssymb,aps,11pt]{revtex4-2}

\usepackage{bm}
\usepackage{amsfonts}
\usepackage{latexsym}
\usepackage[latin1]{inputenc}
\usepackage{graphicx}
\usepackage{amsmath}
\usepackage{palatino}
\usepackage{mathpazo}
\usepackage[normalem]{ulem}
\usepackage{xcolor}

\usepackage{booktabs}
\usepackage{dcolumn}

\usepackage{natbib}

\def\jnl@style{\it}
\def\aaref@jnl#1{{\jnl@style#1}}

\def\aaref@jnl#1{{\jnl@style#1}}

\def\aj{\aaref@jnl{AJ}}                   % Astronomical Journal
\def\apj{\aaref@jnl{ApJ}}                 % Astrophysical Journal
\def\apjl{\aaref@jnl{ApJ}}                % Astrophysical Journal, Letters
\def\apjs{\aaref@jnl{ApJS}}               % Astrophysical Journal, Supplement
\def\apss{\aaref@jnl{Ap\&SS}}             % Astrophysics and Space Science
\def\aap{\aaref@jnl{A\&A}}                % Astronomy and Astrophysics
\def\aapr{\aaref@jnl{A\&A~Rev.}}          % Astronomy and Astrophysics Reviews
\def\aaps{\aaref@jnl{A\&AS}}              % Astronomy and Astrophysics, Supplement
\def\mnras{\aaref@jnl{Mon.~Not.~Roy.~Astron.~Soc.}}             % Monthly Notices of the RAS
\def\prd{\aaref@jnl{Phys.~Rev.~D}}        % Physical Review D
\def\prc{\aaref@jnl{Phys.~Rev.~C}}  % Physical Review C
\def\prl{\aaref@jnl{Phys.~Rev.~Lett.}}    % Physical Review Letters
\def\qjras{\aaref@jnl{QJRAS}}             % Quarterly Journal of the RAS
\def\skytel{\aaref@jnl{S\&T}}             % Sky and Telescope
\def\ssr{\aaref@jnl{Space~Sci.~Rev.}}     % Space Science Reviews
\def\zap{\aaref@jnl{ZAp}}                 % Zeitschrift fuer Astrophysik
\def\nat{\aaref@jnl{Nature}}              % Nature
\def\aplett{\aaref@jnl{Astrophys.~Lett.}} % Astrophysics Letters
\def\apspr{\aaref@jnl{Astrophys.~Space~Phys.~Res.}} % Astrophysics Space Physics Research
\def\physrep{\aaref@jnl{Phys.~Rep.}}      % Physics Reports
\def\physscr{\aaref@jnl{Phys.~Scr}}       % Physica Scripta
\def\commat{\aaref@jnl{Comm.~Math.~Phys.}}              % Communications in Mathematical Physics
\def\science{\aaref@jnl{Science}}               % Science
\def\cqg{\aaref@jnl{Classical Quant.~Grav.}}            % Classical and Quantum Gravity
\def\jpcs{\aaref@jnl{JPCS}}                                     % Journal of Physics Conference Series
\def\ijmpd{\aaref@jnl{Int.~J.~Mod.~Phys.~D}}                    % International Journal of Modern Physics D
\def\grg{\aaref@jnl{Gen.~Relat.~Gravit.}}               % General Relativity and Gravitation
\def\rpp{\aaref@jnl{Rep.~Prog.~Phys.}}          % Reports on Progress in Physics
\def\npa{\aaref@jnl{Nucl.~Phys.~A}}        % Nuclear Physics A
\def\lrr{\aaref@jnl{Living Rev.~Rel.}}                   % Living reviews in relativity
\def\jcap{\aaref@jnl{J.~Cosmology Astropart.~Phys.}}    % Journal of cosmology and astroparticle physics
\def\rmp{\aaref@jnl{Rev.~Mod.~Phys.}}   %Reviews of modern physics

\allowdisplaybreaks[1]
\begin{document}
\title{Differentially rotating scalarized neutron stars with realistic post-merger profile }

\author{Kalin V. Staykov}
\affiliation{Department of Theoretical Physics, Faculty of Physics, Sofia University, Sofia 1164, Bulgaria}

\author{Daniela D. Doneva}
\affiliation{Theoretical Astrophysics, Eberhard Karls University of T\"ubingen, T\"ubingen 72076, Germany}
\affiliation{INRNE - Bulgarian Academy of Sciences, 1784  Sofia, Bulgaria}

\author{Lavinia Heisenberg}
\affiliation{Institut f\"ur Theoretische Physik, Philosophenweg 16, 69120 Heidelberg, Germany}
\affiliation{Institute for Theoretical Physics, ETH Zurich, Wolfgang-Pauli-Strasse 27, 8093 Zurich, Switzerland}
\affiliation{Perimeter Institute, 31 Caroline Street N, Waterloo ON, Canada}

\author{Nikolaos Stergioulas}
\affiliation{Department of Physics, Aristotle University of Thessaloniki, 54124 Thessaloniki, Greece}

\author{Stoytcho S. Yazadjiev}
\affiliation{Theoretical Astrophysics, Eberhard Karls University of T\"ubingen, T\"ubingen 72076, Germany}
\affiliation{Department of Theoretical Physics, Faculty of Physics, Sofia University, Sofia 1164, Bulgaria}
\affiliation{Institute of Mathematics and Informatics, 	Bulgarian Academy of Sciences, 	Acad. G. Bonchev St. 8, Sofia 1113, Bulgaria}

	\begin{abstract}
The merger remnant of a binary neutron star coalescence is initially strongly differentially rotating. Some properties of these remnants can be accurately modeled through building equilibrium neutron star models. In the present paper, we study how a modification of general relativity, namely scalar-tensor theory with a massive scalar field, will alter the picture. In contrast to previous studies, we implement a realistic phenomenological differential rotational law which allows for neutron star models to attain maximal angular velocity away from the center. We find that solutions with much higher masses and angular momenta exist in scalar-tensor theory compared to general relativity. They keep their quasi-spherical energy-density distribution for significantly higher values of the angular momentum before transitioning to quasi-toroidal models, in contrast to pure general relativity. Constructing such neutron star solutions is the first step to our final goal that is studying how scalarization alters the stability and gravitational wave emission of post-merger remnants.
	\end{abstract}

\maketitle

\section{Introduction}

Gravitational waves observations and the prospect for multimessenger astronomy gives us the tools to study nature's laws in their extreme. Among the best candidates are neutron star (NS) merger events, such as GW170817 \cite{LIGOScientific:2017vwq,LIGOScientific:2017ync}, that allow us to study the behaviour of matter in the condition of extreme density and strong gravity. For that purpose, mergers of binary NSs have been modelled in general relativity (GR) with fully nonlinear three-dimensional simulations, see e.g.
\cite{Hotokezaka_etal_2011, Sekiguchi_etal_2011, Bauswein_Janka_2012,  Hotokezaka_etal_2013,Bernuzzi_etal_2014,Dietrich_etal_2015,Radice_etal_2018,DePietri:2018tpx}. Since one of the main goals of gravitational wave observations is to test gravity in the realm of strong gravitational fields (see e.g. \cite{Yunes:2013dva,Berti:2015itd,LIGOScientific:2019fpa,LIGOScientific:2020tif,LISA:2022kgy} and references therein), a natural extension of these results is to include modified gravity effects. Even in the scope of pure GR, though, merger simulations are very challenging and computationally demanding. Hence, extending them for modified theories of gravity is a highly nontrivial task and has been done only in a handful of cases \cite{Barausse:2012da,Shibata:2013pra,East:2022rqi,Sagunski:2017nzb,Bezares:2021yek}. If one concentrates only on studying certain properties of the merger remnant, though, the problem can be simplified considerably. Assuming that after the merger a neutron star is formed instead of a direct collapse to a black hole, it will be initially strongly differentially rotating. It is possible to construct axisymmetric equilibrium models of such differentially rotating NSs, with properties that resemble those of merger remnants \cite{Camelio:2019rsz,Camelio:2020mdi,Iosif:2020iho,Iosif:2021aum}. Such equilibrium models can serve as a tool to gain important insight into the post-merger dynamics, as detailed below.

In order for this to be done, a realistic differential rotation law, which can reproduce the merger remnant rotational profile, is needed. The well-know and widely used $j$-constant differential rotation law \cite{Komatsu:1989zz} is very useful to gain a first insight in the problem and helped understand the phenomenology. It was found that different types of differentially rotating models can exist, including the quasi-toroidal (the co-called type C), the quasi-spherical (type A) and some additional, more exotic types, see \cite{Ansorg:2008pk,Espino:2019ebx,2019EPJA...55..149B}. The stability of such models was examined through numerical simulations, e.g. in \cite{Duez:2006qe,Giacomazzo:2011cv,Weih:2017mcw,Espino:2019xcl,2023arXiv230206007S}.

The $j$-constant law can not model accurately, though, the post-merger remnant. Perhaps the simplest extension that performs better is the 3-parameter law considered in \cite{Bauswein:2017aur,Bozzola:2017qbu}. A more realistic improvement was developed by Uryu et al. in \cite{Uryu:2017obi,Uryu:2017obi} where a four-parameter differential rotation law was proposed, which allows a neutron star model to attain maximal angular velocity away from the center, a generic property of post-merger remnants seen in all simulations. 
In \cite{Iosif:2020iho,Iosif:2021aum} the authors constructed equilibrium sequences of differentially rotating relativistic stars employing this four-parameter rotation law \cite{Uryu:2017obi}, with parameters motivated by simulations of binary merger remnants. Both polytropic and realistic equations of state (EoS) were considered and both quasi-spherical (type A) and quasi-toroidal (type C) models were constructed.

The quasi-equilibrium models of post-merger remnants can have a variety of applications. This includes the interpretation of the post-merger gravitational wave (GW) spectrum \cite{Bauswein:2011tp,Takami:2014tva}, the study of the threshold mass to prompt collapse to a black hole \cite{Bauswein:2017aur,Rezzolla:2017aly,Shibata:2019ctb}, the modeling of processes taking place on longer secular timescales that can not be easily addressed with nonlinear simulations \cite{Doneva:2015jaa}, etc. Most of the studies on this topic, though, are performed in pure GR. Modification of the theory of gravity can significantly influence the neutron star equilibrium and the gravitational wave emission in dynamical processes (e.g. \cite{Blazquez-Salcedo:2018pxo,Blazquez-Salcedo:2022dxh}). An example are the scalar-tensor theories (STT), that are among the most natural and widely studied modification of GR, where the additional scalar field can lead to considerable increase of the stellar maximum mass, especially in the case of rapid rotation \cite{Damour:1993hw,Doneva:2013qva,Doneva:2016xmf}. The quasinormal mode spectrum can be significantly altered as well, e.g. \cite{Sotani:2004rq,Blazquez-Salcedo:2020ibb,Kruger:2021yay} and new types of oscillations can appear, e.g. \cite{2021PhRvD.104j4036M}. All this hints towards the idea that the predictions mentioned above in GR might change significantly in the presence of additional fields. Thus, the gravitational wave observations will allow us to put strong constraints on modified gravity.

The study of quasi-equilibrium rotating post-merger models in modified gravity is a practically undeveloped area. As a matter of fact, there are just a handful of papers constructing uniformly rotating neutron stars in modified gravity \cite{Doneva:2013qva,Kleihaus:2014lba,Doneva:2016xmf,Kleihaus:2016dui}, while differential rotation was considered only in \cite{Doneva:2018ouu}. The differential rotation law employed in the latter paper was the simplified $j$-constant law that can not model well post-merger remnants. It still demonstrated, though, how different post-merger quasi-equilibrium models can differ from GR (for results from fully nonlinear dynamical merger simulations in modified gravity see \cite{Shibata:2013pra}). This motivates us to extend on the above-mentioned papers and study differentially rotating models in scalar-tensor theory with the realistic four-parameter rotation law of \cite{Uryu:2017obi}, which attains maximal angular velocity away from the center. These models can be used as a background for studying  stability and gravitational wave emission that will be done in a follow-up paper. 

Another extension to \cite{Doneva:2018ouu} is that  we consider the cases of nonzero scalar field mass. Apart from the fact that this is an interesting generalization of the model, the scalar field mass has another very important role -- to help us evade the binary pulsar observation that for the moment have ruled out completely scalarization in the classical scalar-tensor theories with zero scalar field potential \cite{Zhao:2022vig}. This property comes from the fact that for a nonzero scalar field mass the scalar field has a finite range of the order of its Compton wavelength. Thus, if this Compton wavelength is below the binary pulsar separation, the two orbiting compact objects would not feel each other's field \cite{Ramazanoglu:2016kul,Yazadjiev:2016pcb}. A scalar field mass as small as $10^{-16}$eV is sufficient to evade all binary pulsar constraints while leaving the equilibrium compact object models and their dynamics almost indistinguishable from the massless scalar field case\footnote{The effect of the scalar field mass starts to be well pronounced in the neutron star structure and dynamics only for a few orders of magnitude larger mass than $10^{-16}$eV.}.

The structure of this paper is as follows: in Section II we briefly present the mathematical background for the considered STT and for the differential rotation. In Section III we present the numerical results. The paper ends with a Conclusion. 

\section{Scalar-tensor theory and differential rotation low }

The general form of the Einstein frame action for scalar-tensor theories  is   
\begin{eqnarray}
S= \frac{1}{16\pi G_{*}}\int d^4x \sqrt{-g} \left(R -
2g^{\mu\nu}\partial_{\mu}\varphi \partial_{\nu}\varphi -
4V(\varphi)\right)+ S_{m}[\Psi_{m}; {\cal A}^{2}(\varphi)g_{\mu\nu}],
\end{eqnarray}
where $R$ is the Ricci  scalar with respect to the Einstein frame metric $g_{\mu\nu}$, $V(\varphi)$ is the scalar field potential and $ {\cal A}^{2}(\varphi)$ is the Einstein frame coupling function between the matter and the scalar field that appears in the action of the matter $S_{m}$. The matter fields are collectively denoted by $\Psi_{m}$. The field equations obtained after varying the action above have the form 
\begin{eqnarray} 
R_{\mu\nu} - \frac{1}{2}g_{\mu\nu}R &=& 8\pi G_{*} T_{\mu\nu}
+ 2\partial_{\mu}\varphi \partial_{\nu}\varphi   -
g_{\mu\nu}g^{\alpha\beta}\partial_{\alpha}\varphi
\partial_{\beta}\varphi -2V(\varphi)g_{\mu\nu}  \,\,\, , \label{EFFE1}\\
\nabla^{\mu}\nabla_{\mu}\varphi &=& - 4\pi G_{*} k(\varphi)T
+ {dV(\varphi)\over d\varphi} , \label{EFFE}
\end{eqnarray}
where $k(\varphi)$ is the coupling function defined as
\begin{equation}
k(\varphi)= \frac{d\ln({\cal  A}(\varphi))} {d\varphi}.
\end{equation}

In the present paper we will consider theory with massive scalar field with potential 
\begin{equation}
    V(\varphi) = \frac{1}{2}m_{\varphi}^2\varphi^2,
\end{equation}
and coupling function 
\begin{equation}\label{eq:coupling}
k(\varphi) = \beta \varphi + \alpha,
\end{equation}
where $\beta$ and $\alpha$ are constants. The case of $\alpha \ne 0$ and $\beta=0$ corresponds to the famous Brans-Dicke theory while for $\alpha = 0$ and $\beta \ne 0$ we have the Damour-Esposito-Farese model allowing for neutron star scalarization \cite{Damour:1993hw}. Interestingly, when $\alpha \neq 0$ the radial perturbations of a compact object also lead to gravitational wave emission related with the presence of the so-called breathing polarisation modes. 

The idea behind introducing a scalar field potential is not only to explore possible modifications beyond GR in their full richness. As discussed in the introduction, even more important is that even a very small $m_{\varphi}$, that practically has a negligible effect on the structure of compact objects, can evade the binary pulsar constraints \cite{Zhao:2022vig} and allow for larger deviations from GR in the strong field regime. 

We consider stationary and axisymmetric matter and scalar field configurations that allow us to use the following form of the metric:
\begin{eqnarray}
&&ds^2 = -e^{\gamma+\sigma} dt^2 + e^{\gamma-\sigma} r^2
\sin^2\theta (d\phi - \omega dt)^2 + e^{2\alpha}(dr^2 + r^2
d\theta^2),
\end{eqnarray}
where all metric functions depend only on  $r$ and $\theta$. The explicit form of the dimensionally reduced field equations is quite lengthy and that is why we refer the interested reader to \cite{Doneva:2013qva, Doneva:2016xmf, Yazadjiev:2015zia}. 

Having discussed the metric and the scalar field, let us focus now on how the matter is treated. The conservation of the energy-momentum tensor in the Einstein frame takes the following form
\begin{eqnarray}
\nabla_{\mu}T^{\mu}{}_{\nu} = k(\varphi)T\partial_{\nu}\varphi .
\end{eqnarray}
We will assume that the NS are made out of a perfect fluid with an energy-momentum tensor being
\begin{eqnarray}
T_{\mu\nu}= (\varepsilon + p)u_{\mu} u_{\nu} + pg_{\mu\nu},
\end{eqnarray}
where $p$ and $\varepsilon$ are the pressure and energy density of the fluid. The Einstein frame fluid four velocity takes the form
\begin{equation}
u^\mu = \frac{e^{-(\sigma + \gamma)/2}}{\sqrt{1-v^2}}
[1,0,0,\Omega], \label{eq:four_velocity}
\end{equation}
where $\Omega=\frac{u^{\phi}}{u^{t}}$ is the angular velocity and the proper velocity $v$ of the fluid is given by $v = (\Omega - \omega) r \sin \theta e^{-\sigma}$. In the differentially rotating case we study here $\Omega = \Omega(r,\theta)$,  while for uniform rotation the angular velocity is a constant.

The quantities defined so far are in the for computations more convenient Einstein frame. The final results presented in this paper, however, are transformed back in the physical Jordan frame. The two frames are related by a conformal transformation and a redefinition of the scalar field. The interested reader can find the detailed relations between the two frames in \cite{Doneva:2013qva}. Here, we will only mention some of the relations we need for the presentation of the results in the paper.

The energy-momentum tensor, the energy density, the pressure and the four velocity transform between the two frames as
\begin{eqnarray}\label{DPTEJF}
T_{\mu\nu} &=& {\cal A}^2(\varphi){\tilde T}_{\mu\nu}, \nonumber \\
\varepsilon &=&{\cal A}^4(\varphi){\tilde\varepsilon}, \nonumber \\
p&=&{\cal A}^4(\varphi){\tilde p},  \\
u_{\mu}&=& {\cal A}^{-1}(\varphi){\tilde u}_{\mu}, \nonumber
\end{eqnarray}
where the Jordan frame quantities are denoted with a tilde. $\Omega$ and $v$ remain the same in both frames. Let us comment on the relation between the Jordan and the Einstein frame mass, radius and angular momentum of the star. The tensor mass of the neutron stars is by definition the ADM mass in the Einstein frame and this mass coincides with the Jordan frame one for the considered coupling function \eqref{eq:coupling}. The angular momentum is by definition the same in both frames as well. The circumferential radius of the star, though, differs and the physical Jordan frame stellar radius is
\begin{equation}
 R_e = {\cal A}(\varphi)\; r \;
e^{(\gamma-\sigma)/2}|_{r=r_{e},\theta=\pi/2},
\end{equation}
where $r_e$ is defined to be the Einstein frame coordinate equatorial radius of the star. $r_{e}$ by itself is defined as the location where the pressure vanishes ${\tilde p}(r_{e},\theta=\pi/2)=0$.

The field equations and the equation for hydrostationary equilibrium should be supplemented with an EoS. The equation of state, however, is given in the physical Jordan frame, hence it is more convenient to use ${\tilde \varepsilon}$ and ${\tilde p}$ for the rest of the section. With the above assumptions and notations, the hydrostationary equilibrium equation has the following explicit form
\begin{eqnarray}
\frac{\partial_i{\tilde p}}{{\tilde \varepsilon} + \tilde{p}} -
\left[\partial_i(\ln \, u^t) - u^t u_\phi \partial_i \Omega -
k(\varphi) \partial_i \varphi\right]=0 \label{eq:Hydrostatic_Equil}.
\end{eqnarray}

In the preset paper we will use a realistic four parameter differential rotation law \cite{Uryu:2017obi}
\begin{equation} \label{Eq:DiffRotLaw}
    \Omega = \Omega_c \frac{1 + \left(\frac{F}{B^2\Omega_c}\right)^p}{1 + \left(\frac{F}{A^2\Omega_c}\right)^{p+q}},
\end{equation}
where $F = u^t u_{\phi}$ is the gravitationally redshifted angular momentum per unit rest mass and enthalpy, while $p, q, A,$ and $B$ are constants. When implementing this differential rotation law  we will follow the procedure developed GR in \cite{Iosif:2020iho} taking into account the needed modification related to the addition of a scalar field. The two of the free parameters should be fixed to $p=1$ and $q=3$, which allows the equation for hydrostationary equilibrium to be cast in a simpler form required by the {\tt RNS} code \cite{Uryu:2017obi,Iosif:2020iho}. Instead of directly using the other two parameters $A$ and $B$, though, we introduce new parameters $\lambda_1$ and $\lambda_2$ which are given as relations between the angular velocity at the center of the star $\Omega_c$, the angular velocity at the equator $\Omega_e$, and the maximal angular velocity $\Omega_{max}$ in the following way
\begin{eqnarray}
&&\lambda_1 = \frac{\Omega_{max}}{\Omega_c},\\
&&\lambda_2 = \frac{\Omega_{e}}{\Omega_c}.
\end{eqnarray}
When one sets values for $\lambda_1$ and $\lambda_2$  and substitute the rotational law (\ref{Eq:DiffRotLaw}) in the above relations, the obtained system is solved for $A$ and $B$ at each iteration. The motivation behind the use of $\lambda_1$ and $\lambda_2$ roots in the fact that it is easier to reproduce the Newtonian limit \cite{Uryu:2017obi} and it is more convenient for controlling the type of differentially rotating model, e.g. whether it is quasi-toroidal or quasi-spherical. This problem is discussed in greater detail in \cite{Iosif:2020iho} and we refer the interested reader to this paper for further information.

\section{Numerical results}
The main goal of the present paper is to study the effect of the scalar-tensor theory on differentially rotating equilibrium neutron stars that can serve as a model  of post-merger remnants  with the more realistic four-parameter differential rotation law (\ref{Eq:DiffRotLaw}). Hence, it would be enough to concentrate on a single realistic cold equation of state and we have chosen the MPA1 EOS \cite{Muther:1987xaa}. Even with the single EoS, and fixing $p$ and $q$ in \eqref{Eq:DiffRotLaw}, we  are left with five free parameters, namely three connected to the STT ($\beta$, $\alpha$, $m_{\varphi}$), and two connected with the differential rotation ($\lambda_1$, $\lambda_2$). Since the focus of the paper is on the modified gravity effects, and in order to keep the outline consize, we have decided to fix $\lambda_1$ and $\lambda_2$ and concentrate on the effect of changing the STT parameters ($\beta$, $\alpha$, $m_{\varphi}$). The exploration of the former parameters, especially in connection to actual merger simulation in STT, will be done in a follow-up publication. 

We will focus on a combination of ($\lambda_1$, $\lambda_2$) that leads to quasi-toroidal models in GR. In STT, though, the quasi-toroidal energy-density distribution appears only for extremely large angular momenta. The simulations of binary neutron star remnants, on the other hand, show that such quasi-toroidal models are a transient phase appearing right after the merger. Shortly afterwards, the remnants settle into type A, quasi-spherical, models. Thus the fact that producing a quasi-toroidal model in the presence of a strong enough scalar field is possible only in a very limited region of the parameter space, can have interesting astrophysical consequences.

Based on previous studies in GR \cite{Iosif:2020iho, Iosif:2021aum} we have chosen to work with ${\lambda_1=1.5,\lambda_2=0.5}$. For the numerical calculations we developed a modification of the {\tt RNS} code \cite{Stergioulas:1994ea,Doneva:2018ouu}. All quantities presented in this section are in the physical Jordan frame.

\subsection{General properties for sequences of diffentially rotating neutron stars} 
\begin{figure}[]
	\centering
	\includegraphics[width=0.44\textwidth]{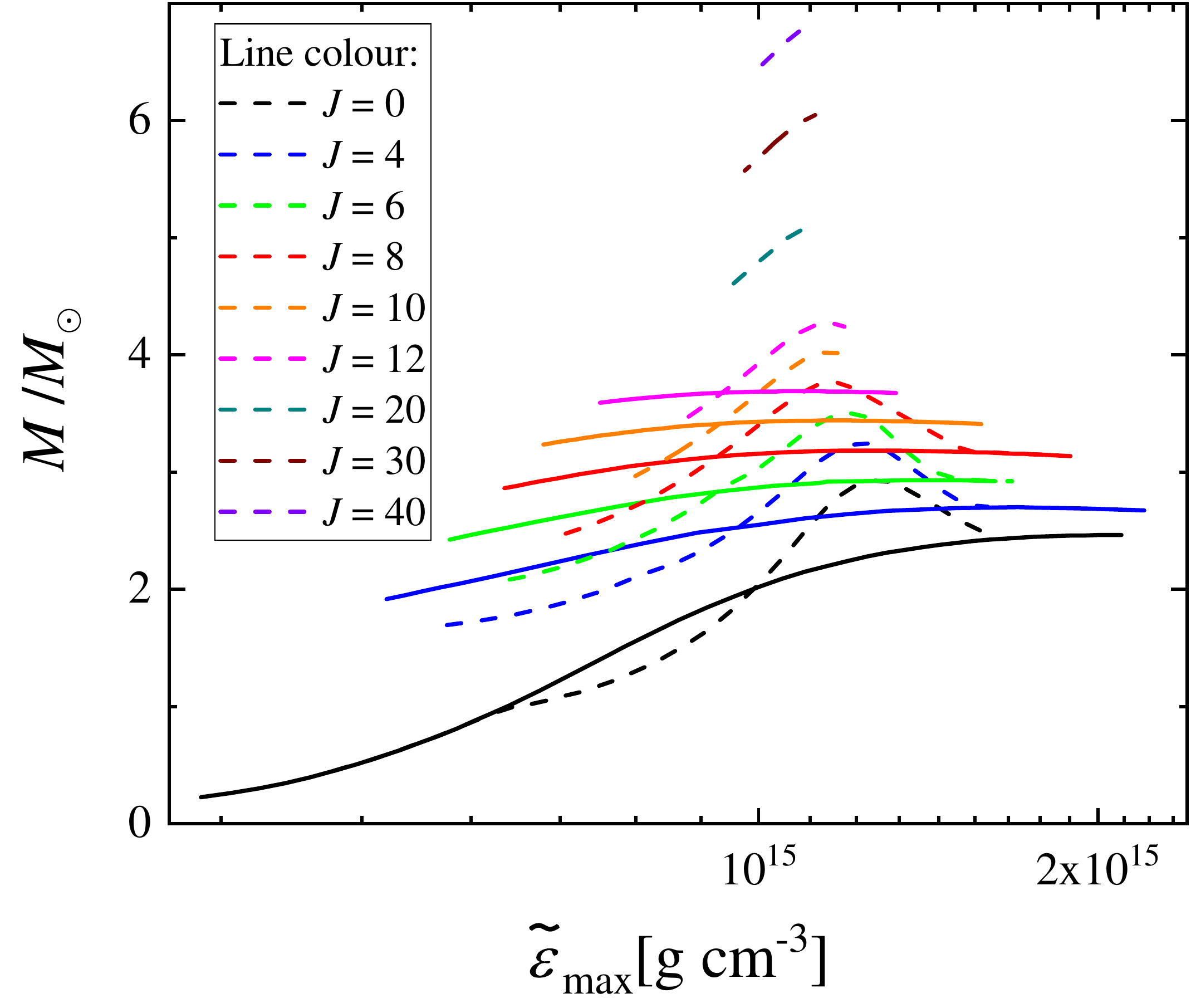}
	\includegraphics[width=0.45\textwidth]{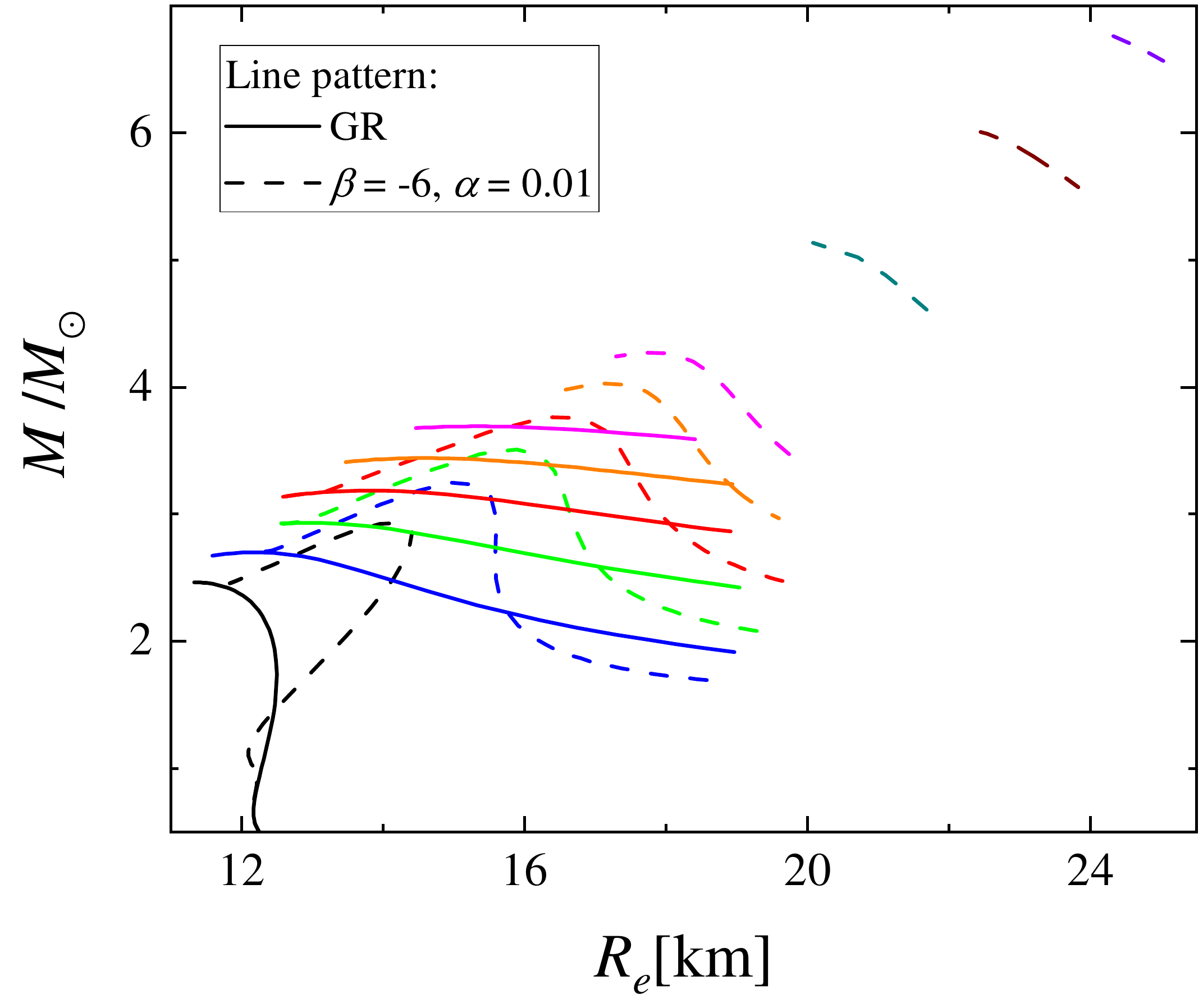}
	\caption{\textit{Left:} The neutron star mass as function of its maximal energy density \textit{Right:} The neutron star mass as a function of its equatorial radius. The presented models are for $J$ (in geometrical units: $J = \frac{J c}{G M^2_{\odot}}$)  spanning from $J=0$ to values close to the highest possible we could calculate in GR and STT. GR is  depicted with continuous lines while the STT models are plotted with in dashed lines. The different values of the angular momentum are in different colours. }
	\label{Fig:M_R_eps}
\end{figure}

As a first step in our study we concentrate on the global properties of the differentially rotating models in STT, and how the parameters of the theory and the mass of the field effect them. For that purpose it will be most suitable to examine sequences of models with constant angular momentum $J$ (in dimensionless units $J = \frac{J c}{G M^2_{\odot}}$) similar to the studies in GR \cite{Iosif:2020iho,Iosif:2021aum} since this is one of the conserved quantities.  In Fig. \ref{Fig:M_R_eps} we present the mass of the star as a function of the maximal energy density in the left panel and as a function of the equatorial radius in the right panel. The presented sequences of models are for GR and  massless STT with some representative values of the parameters $\beta = -6$, $\alpha = 0.01$ and $m_\varphi=0$. As we will see below, these results are practically indistinguishable from the case of $m_\varphi=10^{-16}$eV that is roughly the minimum one allowing to safely evade binary pulsar observations.

Let us first focus on the nonrotating case with $J=0$. Strictly speaking, for $\alpha \neq 0$ all models are endowed with scalar field. Thus, GR is not a solution of the field equations and spontaneous scalarization in the sense of \cite{Damour:1993hw} is not possible. For the chosen parameters in the STT, though, we observe a scenario which closely resembles scalarization. For low compactness the GR and the STT branches practically overlap since the scalar field is negligible. After a certain critical compactness is reached the scalar field sharply increases that resembles the onset of scalarization and the first bifurcation-like point is formed.  
This scalarized branch reaches a maximum mass and afterwards the scalar field decreases again to very small value and practically merges again with the GR one at a second bifurcation-like point.

After discussing the $J=0$ sequences let us move to larger $J$ in Fig. \ref{Fig:M_R_eps}. In the figure we start with $J=4$ (in geometrical units) that corresponds already to a very fast rotation. As seen, we could not obtain the first bifurcation-like point numerically but at least the scalarized sequence have a well pronounced shape with a maximum and second bifurcation-like point appearing. With the increase of $J$ the sequences get increasingly short and for very large $J$ we could obtain only a small part of the neutron star sequence, both for GR and STT. The problem of generating longer sequences of solutions is numerical as discussed in detail in \cite{Iosif:2020iho,Iosif:2021aum}, namely the modified {\tt RNS} code fails to converge to a unique solution.  Nevertheless, we invested a lot of efforts in obtaining long enough sequences of deferentially rotating neutron stars so that we can judge about the new physics induced by the presence of a scalar field.  

The first important observation is that in GR we were able to find solutions with $J$ up to roughly $J = 12$ as seen in Fig. \ref{Fig:M_R_eps}. In STT with $\beta = -6$ and $\alpha = 0.01$, however, the situation is dramatically different and solutions with much higher $J$ exist. We managed to reach up to roughly $J = 50$ but we have indications that higher $J$ might be also possible. Similar indications for larger $J$ sequences in STT were already observed for neutron stars with $j$-constant differential rotation law \cite{Doneva:2018ouu} but the maximum $J$ that was reached there was much smaller. As expected, with the increase of $J$, we observe significant increase of the neutron star mass. Such high $J$ models can have very interesting astrophysical implications for the post-merger evolution that will be discussed in the conclusions.

    \begin{figure}[]
	\centering
	\includegraphics[width=0.45\textwidth]{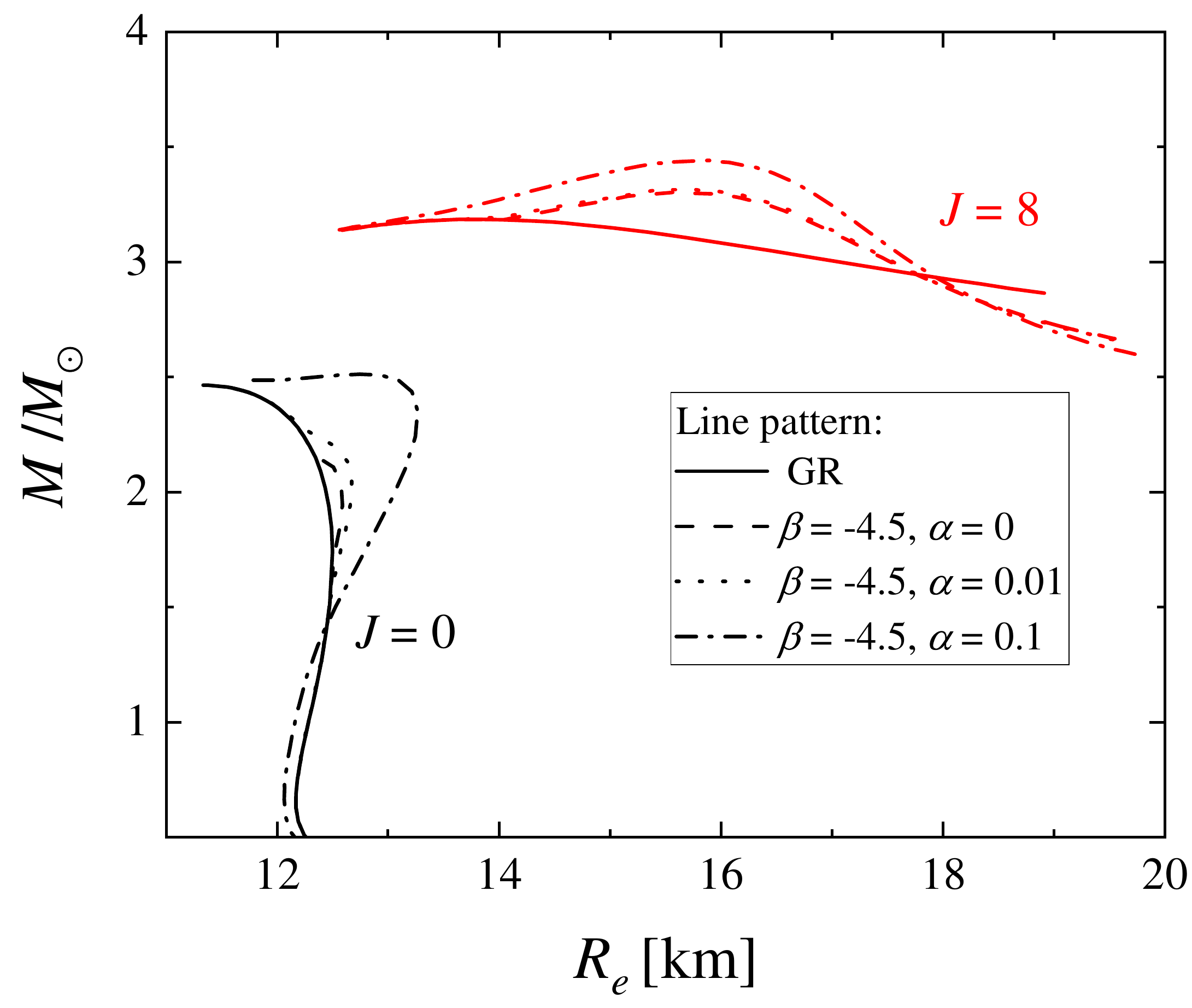}
	\includegraphics[width=0.45\textwidth]{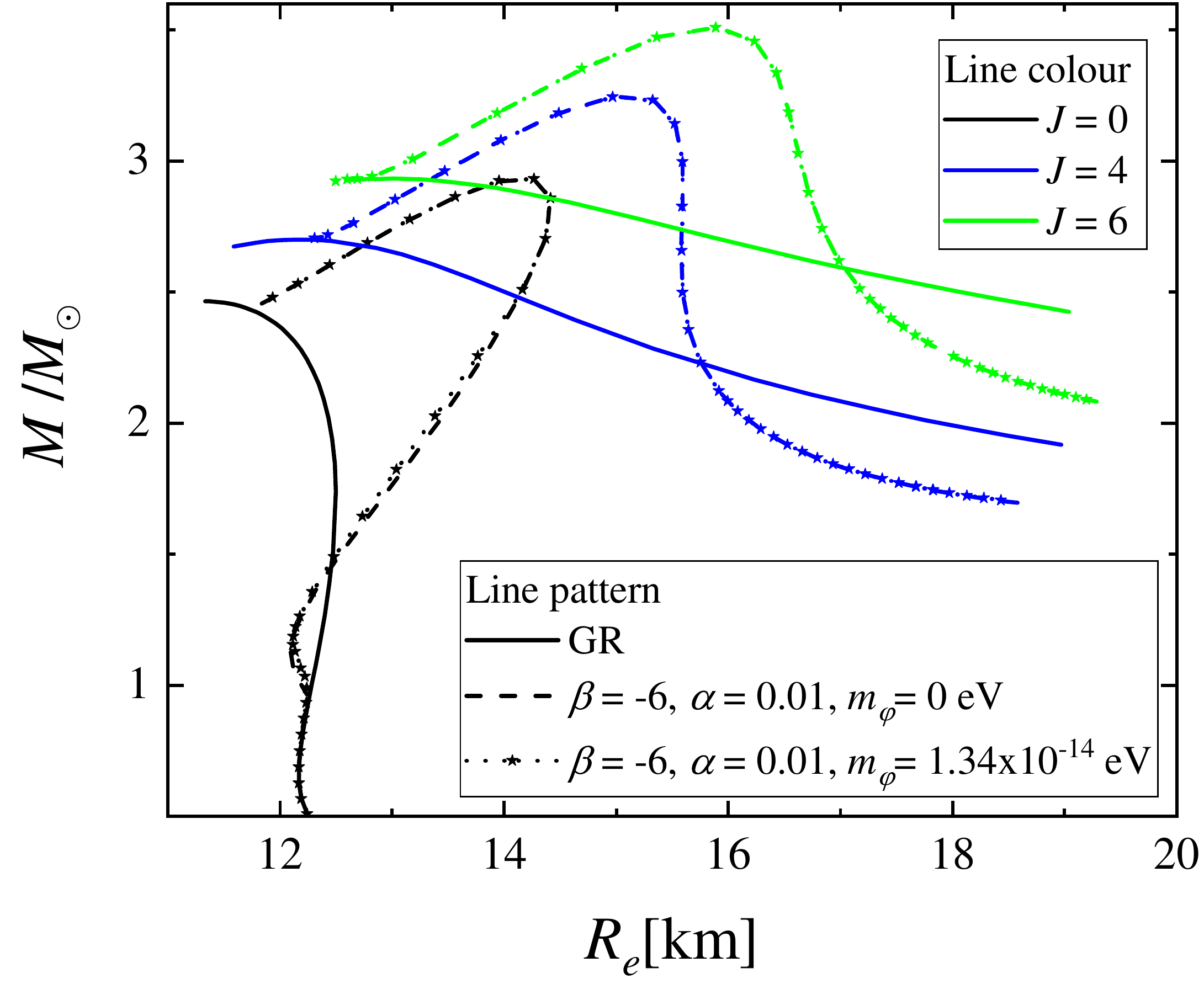}
	\caption{The neutron star mass as function of its equatorial radius. \textit{Left:} Different combinations of ($\beta$, $\alpha$) for non-rotating models and models with moderate values for $J$. \textit{Right:} Non-rotating models and models with two different values for $J$ for fixed STT with nonzero mass $m_{\varphi}>0$. The angular momentum is in geometrical units $\left( J = \frac{J c}{G M^2_{\odot}}\right)$. GR is depicted with continuous lines while the different cases for the STT are  plotted in different line patterns and symbols. The colour coding of the angular momentum is the same as in Fig. \ref{Fig:M_R_eps}.} 
	\label{Fig:M_R_param}
\end{figure}
In Fig. \ref{Fig:M_R_param} we focus on the behavior of the solutions with varying theory parameters ($\beta, \alpha, m_\varphi$). In the left panel we present the mass of the star as a function of its radius for fixed $\beta=-4.5$ and varying $\alpha$ (again for $m_\varphi=0$). The chosen value of $\beta$ is relatively small (in absolute value) in order to demonstrate better the effect of $\alpha$. The presented models are with two fixed angular momenta for which both GR and scalarized solutions could be found, namely $J=0$ and $J=8$. As expected, the increase of $\alpha$ leads to a stronger scalar field and thus deviations from GR but the qualitative picture remains similar. The only major difference is that for larger enough $\alpha$ there are no bifurcation-like points anymore and the neutron star solutions always deviate non-negligibly from GR.

The right panel of Fig. \ref{Fig:M_R_param} serves as a proof of the conjecture expressed above, namely that a small scalar field mass that can evade binary pulsar experiments, have no measurable influence on the neutron star structure with respect to the massless case. In order to demonstrate that we present the mass of the star as a function of its radius for fixed  $\beta = -6$ and $\alpha = 0.01$, and two masses, $m_\varphi=0$ and $m_{\varphi} = 1.34\times10^{-14}$eV. As seen, there is no visible difference between the massive and the massless case. This justifies the fact that for simplicity and convenience from a numerical point of view, a big portion of our results are presented for $m_\varphi=0$.

\subsection{Profiles of scalarized differentially rotating neutron stars}
    \begin{figure}[]
	\centering
	\includegraphics[width=0.95\textwidth]{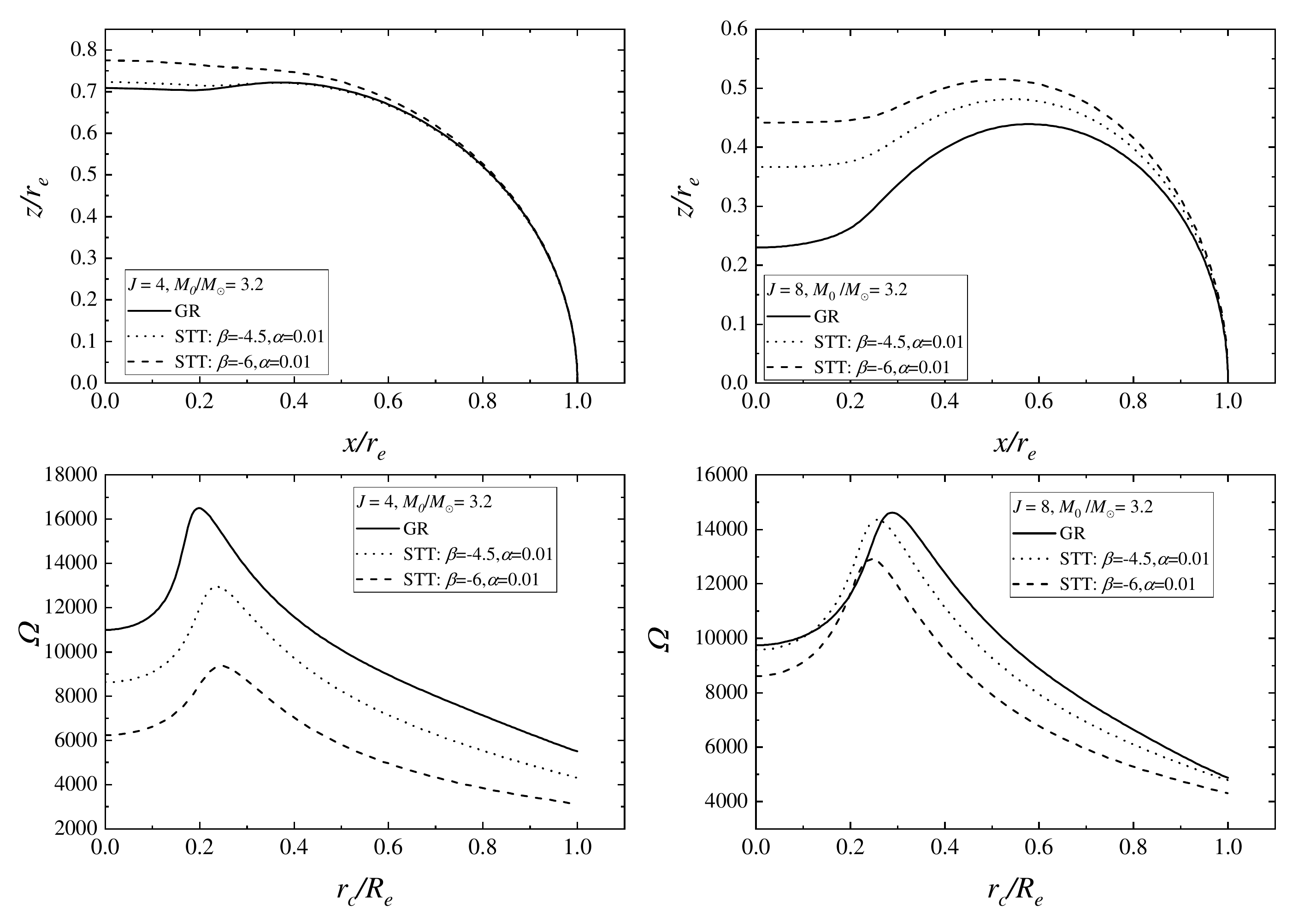}
	\caption{Models with low and moderate values for $J$ (in geometrical units), fixed baryon mass $M = 3.2 M_{\odot}$ and two cases of STT. \textit{Top row:} Surface of the star in the $x-z$ plane, where both axes are normalized to the equatorial coordinate radius $r_e$. \textit{Bottom row:} The equatorial distribution of the angular velocity where the circumferential radial coordinate $r_{c}$ is normalized to the circumferential radius of the star $R_e$. }
	\label{Fig:GR_STT}
\end{figure}
We continue our study of  differentially rotating scalarized neutron stars by examining individual models.
In Fig. \ref{Fig:GR_STT} we compare the surface profiles of the stars, plotted in the $x-z$ plane (both coordinates are normalized to the equatorial coordinate radius $r_e$) (top row) and the angular velocity distributions in the equatorial plane as a function of the circumferential radial coordinate $r_{c}$, normalized to the circumferential radius of the star $R_e$ (bottom row). All models have fixed baryon mass $M_0 = 3.2 M_{\odot}$ and two different values of the angular momentum, namely $J = 4$ and $J = 8$. The parameter $\alpha$ is fixed to  $\alpha = 0.01$ and two different values for $\beta$ are presented. As one can see, for the same $J$, the GR models are the most toroidal ones while with the decrease of $\beta$ and thus the increase of the scalar field strength, the toroidal shape is softened. As a general observation we can say that in the STT, the differentially rotating models stay spherical (or quasi-spherical) for much higher values of $J$ (even beyond those allowed in GR). This was also observed for differentially rotating models with $j$-constant law \cite{Doneva:2018ouu}. Concerning the angular velocity, in the GR case it is the highest and it decreases with the decrease of $\beta$.

    \begin{figure}[]
	\centering
	\includegraphics[width=0.95\textwidth]{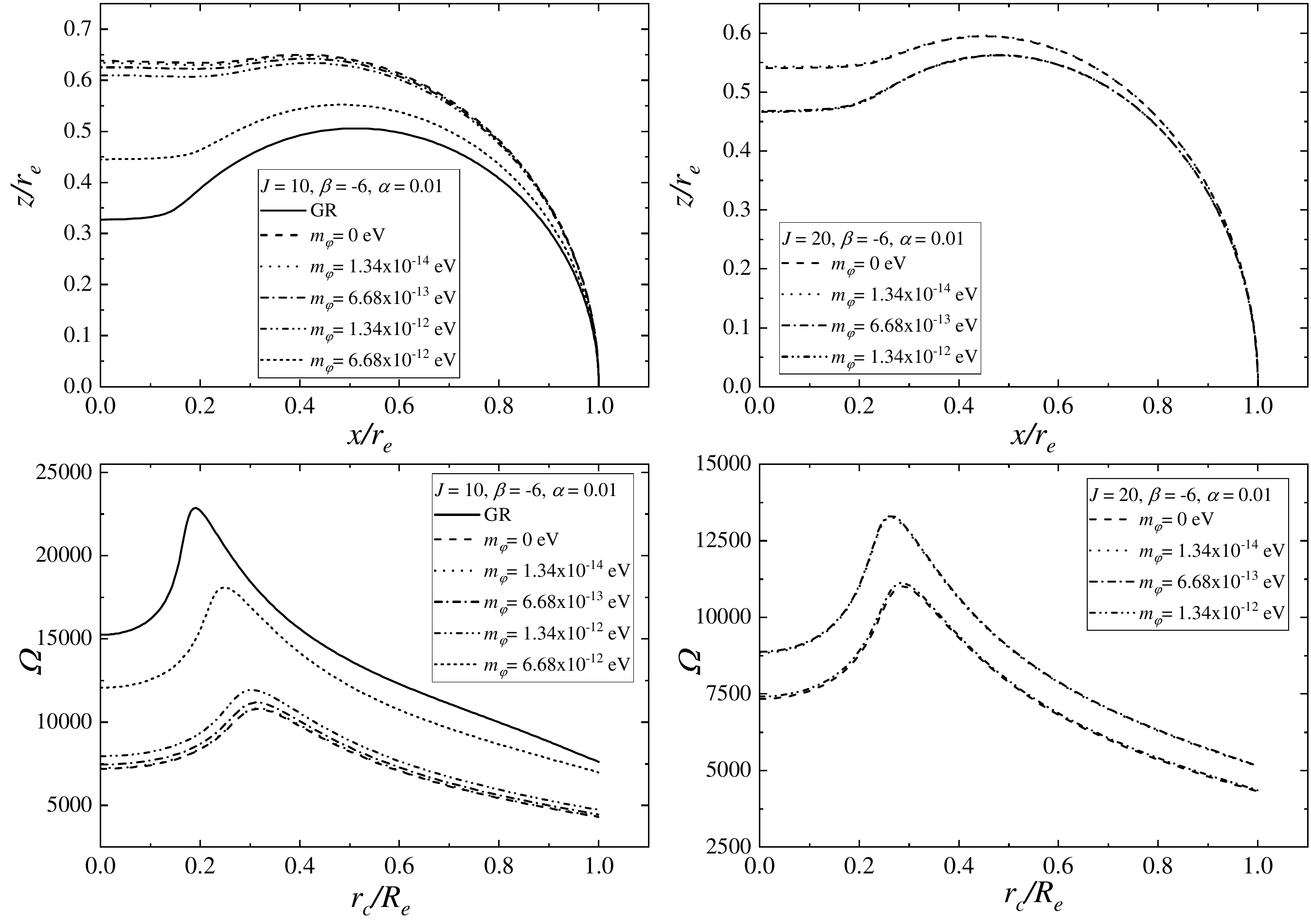}
	\caption{Models with $J=10$ \textit{(left column)} and $J=20$ \textit{(right column)}, fixed parameters of the STT and different values for the scalar field mass $m_{\varphi}$. All models are with fixed maximal energy density $\tilde{\varepsilon}_{\rm max} = 1\times 10^{15} {\rm g\;cm^{-3}}$. \textit{Top row:} surface of the star in the $x-z$ plane, where both axes are normalized to the equatorial coordinate radius $r_e$. \textit{Bottom row:} the equatorial distribution of the angular velocity  where the circumferential radial coordinate $r_{c}$ is normalized to the circumferential radius of the star $R_e$. The GR case (if exists) is depicted with continuous lines and the different values for the mass of the scalar field are plotted with different patterns.}  
	\label{Fig:massive}
\end{figure}

In Fig. \ref{Fig:massive} we study the effect of the scalar field mass on individual models as well as the effect of extremely high $J$ for which only scalarized solutions exist. Thus, we present two values of $J$, namely $J=10$ and $J=20$, and several scalar field masses, while  $\beta = -6$ and $\alpha = 0.01$ are kept fixed. The models we study are with fixed maximal energy-density $\tilde{\varepsilon}_{\rm max} = 1\times10^{15} {\rm g\;cm^{-3}}$. The reason is that this is more convenient than a fixed baryon mass in the case of high $J$ where neutron star solutions exist in a much more narrow range of parameters.
As one can expect, based on the known results for massive STT \cite{Yazadjiev:2016pcb,Doneva:2016xmf}, the increase of the scalar field mass suppresses the effect of scalarization and the results converge to the GR case (if it exist) as the mass of the field goes to infinity.  Therefore, the models get more toroidal with the increase of $m_\varphi$ and the angular velocity increases as well as seen in Fig. \ref{Fig:massive}. For all considered combinations of $J$ and $m_\varphi$, though, the toroidal deformation remains still moderate in comparison with GR.

\begin{figure}[]
	\centering
	\includegraphics[width=0.95\textwidth]{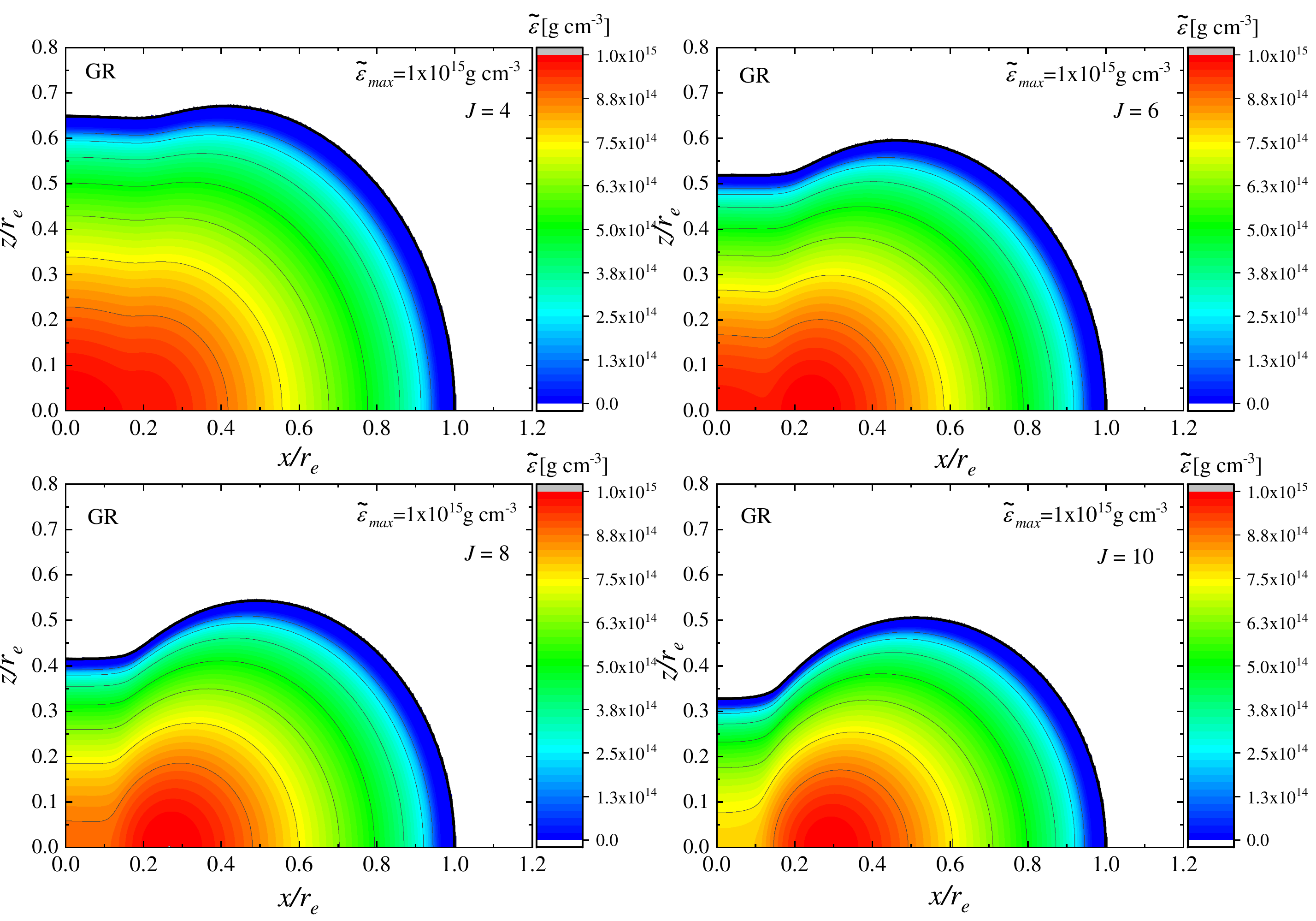}
	\caption{Contour plots of the energy-density distribution for models with different values for the angular momentum (in geometrical units) in GR. The energy-density is in ${\rm g\;cm^{-3}}$. The maximal energy density is fixed for all models, $\tilde{\varepsilon}_{\rm max} = 1\times10^{15} {\rm g\;cm^{-3}}$, and it the same as in Fig. \ref{Fig:STT_cplot}. }
	\label{Fig:GR_cplot}
\end{figure}

The reason why we observe less-toroidal profiles for larger absolute values of $\beta$ lies in the geometry of the scalar field. As already observed  even for uniformly rotating models \cite{Doneva:2013qva}, despite the larger deformation of the neutron star fluid shape away from a sphere in the case of very rapid (differential) rotation, the geometry of the scalar field remains nearly spherical. In order to better understand the effect of the scalar field in our case, we proceed with examining contour plots of the energy density and the scalar field distribution. In the STT case we span from $J=10$ up to $J=40$ and we plot models with the same maximal energy density  $\tilde{\varepsilon}_{\rm max} = 1\times10^{15} {\rm g\;cm^{-3}}$. Before we proceed to the STT case, though, as a point of reference, in Fig. \ref{Fig:GR_cplot} we present the contour plots of the energy density distribution in GR for values of $J$ in the range from $J = 4$ to $J = 10$ (around the maximum one we could achieve in GR).  In this case as well, the maximal energy density is the same for all models (for consistency with the STT case), and it is the same as the one used in the STT case which follows.

\begin{figure}[]
	\centering
	\includegraphics[width=0.95\textwidth]{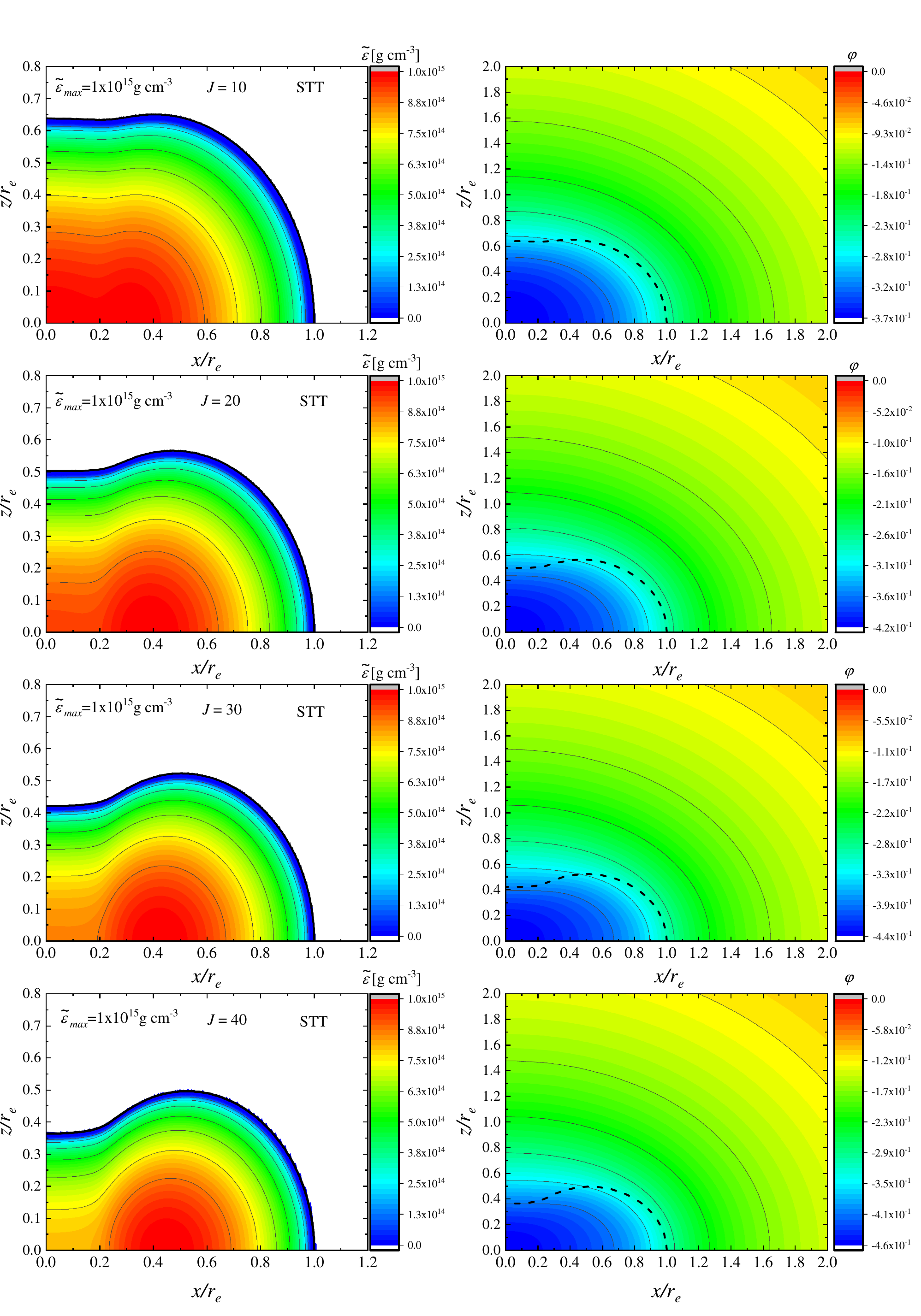}
	\caption{Contour plots for scalarized neutron stars models with high angular momentum. The STT parameters are fixed to $\beta = -6$, $\alpha = 0.01$ and $m_{\varphi} = 0$, and the maximal energy density is the same for all models: $\tilde{\varepsilon}_{\rm max} = 1\times10^{15} {\rm g\;cm^{-3}}$. \textit{Left:} The energy-density distribution measured in ${\rm g\;cm^{-3}}$. \textit{Right:} The scalar field distribution where the stellar surface is depicted given with dashed line on top of the scalar field contour plot.  }
	\label{Fig:STT_cplot}
\end{figure}

In Fig. \ref{Fig:STT_cplot} we present the contour plots of the energy density (the left column) and the scalar field (right column) in the massless STT with $\beta = -6$ and $\alpha = 0.01$. The behaviour of the energy density distribution is qualitatively similar in GR and STT -- the maximal energy density shifts away from the center with the increase of $J$. It is interesting to note that this behaviour is significantly delayed towards larger $J$ in STT compared to GR. In addition, we observe here again that the fluid distribution of the  GR models have more pronounced quasi-toroidal structure, while in the STT case, the models are flatter, though, the quasi-toroidal structure is present as well. The reason for this behaviour we can potentially find in the scalar field distribution. With the increase of $J$ the scalar field gets more oblate. The maximum of the scalar field, though, does not shift from the center of the star. This is the reason why STT models can support so high values for the angular momentum, compared with the GR case.  

\section{Conclusion}

A rapidly and strongly differentially rotating neutron star can be born as an outcome of a binary neutron star merger.  The merger remnant can exist for a few tens of milliseconds supported by the differential rotation before collapsing to a black hole or relaxing to a rigidly rotating neutron star in equilibrium. This final stage of the merger has the potential to test the behaviour of matter in the regime of strong gravitational fields and high energy-density, as well as the underlying gravitational theory. Even though these remnants are produced in a highly dynamical scenario, an approximate but accurate method to explore them is to build quasi-equilibrium models of neutron stars using a phenomenological differential rotational law matched to the results from the nonlinear dynamical simulations.

In the present paper, we studied differentially rotating neutron star models in massive scalar-tensor theory using Uryu et al.'s realistic phenomenological  4-parameter rotation law. The models were computed with a modification of the {\tt RNS} code. We investigated a broad range of theory parameters and the scalar field coupling function we have chosen allows for the presence of the so-called breading modes -- an extra gravitational wave polarization potentially detectable by the ground-based instruments. The addition of a scalar field mass, on the other hand, allows us to circumvent the binary pulsar constraints, that are otherwise very restrictive in the massless scalar field case. This widens the range of parameters to be studied and opens up a possibility for larger deviations from GR.

Contrary to previous studies in STT, the realistic differential rotation law we consider allows the maximum of the angular velocity to be shifted away from the stellar center which is in agreement with the rotational profiles of the remnants produced in merger simulations. Thus, Uryu et al.'s law can produce both quasi-toroidal (type C) and quasi-spherical (type A) axisymmetric configurations which can more or less accurately model different phases of the merger remnant evolution. The results show that the scalarized neutron stars can reach significantly higher angular momenta that are a few times the maximum ones in GR. To be more precise, in GR we found solutions with maximal angular momentum about  $J = 13$ (in geometrical units) while in the STT case, we were able to reach beyond $J = 50$. In addition, we found that in STT a neutron star stays quasi-spherical for values of the angular momentum which are well above the maximal one allowed in GR, and the models became quasi-toroidal only for extremely high values of $J$. The reason behind this is that even for extremely rapid rotation the scalar field remains quasi-spherical. The indirect influence of the scalar field on the density and angular velocity distribution of the fluid leads to less quasi-toroidal STT models compared to GR. It will be very interesting to confront these findings against results from merger simulations to see whether the actual merger remnants are also affected in a similar way by the presence of a scalar field, i.e. up to what extent the presence of scalar field can soften the quasi-toroidal distribution and transform it into quasi-spherical. This can potentially have far-reaching implications such as affecting the structure of the kilonova ejecta and making it more spherical as well. This is actually supported by observations indicating that the kilonova associated with GW170817 has a higher degree of spherical symmetry than what is expected by the present GR simulations \cite{Sneppen:2023vkk}.

The mass of the scalar field on the other hand, as expected, suppresses the scalar field and with the increase of the mass, the results converge to the GR case. It is interesting to point out, however, that for low masses of the field, but high enough to evade the binary pulsar constraints, the results in the massive case coincide with the massless case. This motivates studying the massless case as a limit for maximum deviation from GR that can be achieved even though strictly speaking the theory with $m_\varphi=0$ is excluded from binary pulsar observations.

The results presented in the paper open interesting and important questions. The first one is whether the constructed models, especially with extremely high angular momentum, are stable. The second one is if we can produce such models in actual binary merger simulations. This is a study underway. In any case, our results already give a hint that the standard picture of the post-merger evolution in GR can be significantly altered in STT. This would allow us to test STT in a parameter regime inaccessible both by the binary pulsars and the gravitational wave observations of the inspiral.

\section*{Acknowledgement}
This study is in part financed by the European Union-NextGenerationEU, through the National Recovery and Resilience Plan of the Republic of Bulgaria,
project No. BG-RRP-2.004-0008-C01. DD acknowledges financial support via an Emmy Noether Research Group funded by the German Research Foundation (DFG) under grant no. DO 1771/1-1.
LH is supported by funding from the European Research Council (ERC) under the European Unions Horizon
2020 research and innovation programme grant agreement No 801781 and by the Swiss National Science Foundation
grant 179740. LH further acknowledges support from the Deutsche Forschungsgemeinschaft (DFG, German Research Foundation) under Germany's Excellence Strategy EXC 2181/1 - 390900948 (the Heidelberg STRUCTURES Excellence Cluster). NS is grateful for resources provided by Virgo, which is funded, through the European Gravitational Observatory (EGO), by the French
Centre National de Recherche Scientifique (CNRS), the
Italian Istituto Nazionale di Fisica Nucleare (INFN)
and the Dutch Nikhef, with contributions by institutions from Belgium, Germany, Greece, Hungary, Ireland, Japan, Monaco, Poland, Portugal, and Spain. NS is grateful for networking support through
the COST Action CA18108 "QG-MM". We acknowledge Discoverer PetaSC and EuroHPC JU for awarding this project access to Discoverer supercomputer resources.

\bibliography{references}

\end{document}